\documentclass[sigconf]{acmart}

\usepackage{tikz}
\usepackage{pgfplots}
\pgfplotsset{compat=1.18}
\usepgfplotslibrary{polar}
\usepackage{makecell}
\usepackage{enumitem}

\AtBeginDocument{%
  }

\setcopyright{acmlicensed}
\copyrightyear{2025}
\acmYear{2025}
\setcopyright{rightsretained}
\acmConference[IDC '25]{Interaction Design and Children}{June 23--26, 2025}{Reykjavik, Iceland}
\acmBooktitle{Interaction Design and Children (IDC '25), June 23--26, 2025, Reykjavik, Iceland}
\acmDOI{10.1145/3713043.3731527}
\acmISBN{979-8-4007-1473-3/2025/06}

\usetikzlibrary{shapes.geometric, arrows, positioning}

\tikzstyle{startstop} = [rectangle, rounded corners, minimum width=3cm, minimum height=1cm,text centered, draw=black, fill=red!30]
\tikzstyle{process} = [rectangle, minimum width=3cm, minimum height=1cm, text centered, draw=black, fill=blue!30]
\tikzstyle{decision} = [diamond, minimum width=3cm, minimum height=1cm, text centered, draw=black, fill=green!30]
\tikzstyle{arrow} = [thick,->,>=stealth]
\tikzstyle{data} = [trapezium, trapezium left angle=70, trapezium right angle=110, minimum width=3cm, minimum height=1cm, text centered, draw=black, fill=yellow!30]

\begin{document}

\title{Interaction Analysis by Humans and AI: A Comparative Perspective}


\author{Maryam Teimouri}
\affiliation{%
  \institution{TurkuNLP, University of Turku}
  \city{Turku}
  \country{Finland}}
\email{mtebad@utu.fi}

\author{Filip Ginter}
\affiliation{%
  \institution{TurkuNLP, University of Turku}
  \city{Turku}
  \country{Finland}}
\email{figint@utu.fi}

\author{Tomi "bgt" Suovuo}
\affiliation{%
  \institution{University of Turku} 
  \city{Turku}
  \country{Finland}}
\email{bgt@utu.fi}


\begin{abstract}
This paper explores how Mixed Reality (MR) and 2D video conferencing influence children’s communication during a gesture-based guessing game. Finnish-speaking participants engaged in a short collaborative task using two different setups: Microsoft HoloLens MR and Zoom. Audio-video recordings were transcribed and analyzed using Large Language Models (LLMs), enabling iterative correction, translation, and annotation. Despite limitations in annotations' accuracy and agreement, automated approaches significantly reduced processing time and allowed non-Finnish-speaking researchers to participate in data analysis. Evaluations highlight both the efficiency and constraints of LLM-based analyses for capturing children’s interactions across these platforms. Initial findings indicate that MR fosters richer interaction, evidenced by higher emotional expression during annotation, and heightened engagement, while Zoom offers simplicity and accessibility. This study underscores the potential of MR to enhance collaborative learning experiences for children in distributed settings.
\end{abstract}



\maketitle

\section{Introduction}
Advancements in digital communication—particularly through Mixed Reality (MR) \cite{milgram1994taxonomy} and video conferencing tools like Zoom \cite{zoom}—have reshaped how younger generations interact in educational contexts. This study is part of a larger four-year project funded by the Research Council of Finland, which focuses on developing a HoloLens-based holoportation tool for distributed education. We examine children’s verbal communication during gameplay on an MR platform, with the same experiment replicated on Zoom for comparison. Our analysis draws from both the gameplay sessions and post-experiment interviews. However, challenges such as the time-intensive nature of manual transcription \cite{incollection} and the language barrier posed by Finnish data require extensive data processing. To address these issues, we employ large language models (LLMs) \cite{vaswani2017attention} to support transcription \cite{graves2013speech}, correction \cite{zhang2019recurrent}, translation \cite{johnson2017google}, speaker identification \cite{ma2017text}, and emotion detection \cite{huang2019text}, assessing their effectiveness and practical utility throughout the workflow.


\section{Related Work}
HoloLens creates a Mixed Reality (MR) environment by projecting holograms through depth images captured by Intel RealSense cameras, whose RGBD systems are detailed in \cite{keselman2017intel}, covering optics, noise handling, and performance in gesture recognition and 3D reconstruction. A review of 44 studies from 2016 to 2020 highlights HoloLens’s role in engineering, architecture, and medical education, particularly in enhancing visualization and workflows \cite{app11167259}. Immersive experiences like the Van Gogh exhibition have also been evaluated for their cognitive and emotional impact using sentiment and emotion recognition tools \cite{africa2023salient}, supporting the value of emotion analysis in education, where promoting positive emotional states can enhance engagement and improve learning outcomes \cite{munezero2013exploiting}. Achieving this requires accurate transcription and emotion detection, which recent advances in Large Language Models (LLMs) have significantly improved. Self-supervised models such as wav2vec 2.0 enable better transcription in low-resource languages like Finnish \cite{baevski2020wav2vec}, while speaker identification benefits from approaches like Contextual Beam Search, which combines LLMs and acoustic diarization to enhance accuracy through audio and lexical cues \cite{fritsch2023enhancing}. Recent work also highlights the value of involving children directly in the design of MR experiences. A participatory study with HoloLens 2 and 3D holograms showed how students co-designed social interactions, supporting early integration of educational goals in XR development \cite{goagoses2024primary}. This aligns with the current study’s focus on child-centered, emotionally rich MR environments.

\section{Study Setup}

The study compares two setups: the \textbf{experimental group (A–B)}, where one participant uses a HoloLens to view their partner as a hologram (room A) and the other uses a 3D camera and TV screen (room B); and the \textbf{control group (B–B)}, where both participants used room B, interacting via TV screens and shared digital whiteboards. Rooms are connected with custom software, with HoloLens and 2D cameras in room A, and 3D cameras and TV screens in room B. Tablets on stands allowed drawing on digital whiteboards. Participants received training to become familiar with the devices and interaction style, emphasizing natural communication over technology testing. Familiar images of animals or characters were prepared in advance. In pairs, participants take turns describing and miming gestures of assigned images while their partner guesses and draws on the whiteboard. After five minutes, roles are switched with a new image set. The experimental group experiences both holographic and real-time TV screen interactions, enhancing spatial awareness through MR, while the control group engages only via 2D TV screen setups for comparison.

\section{Data Collection, Data Processing and Tools}

Video and audio feeds are recorded using cameras, HoloLens headsets, and microphones. Lidar cameras are used in 2D mode to prevent interference with HoloLens. For the control group, interactions are recorded via Zoom. Surveys and interviews conducted post-study focus on participants' feelings during the experiment, their familiarity with similar experiences, and their perception of time passing. All recordings are stored securely, and video files were edited and converted to audio for natural language processing tasks.

The original study materials are securely stored on the University's Seafile service (a cloud-based file synchronization and sharing platform), with access restricted to the researchers involved in the study. These materials will be deleted after the study ends, or within a maximum of 5 years. The data will not be shared externally and will remain within Europe.

\textbf{Initial Data:} The dataset includes 10-minute videos from 5 pairs, with 3 pairs in the control group. Videos for the experimental group are recorded separately, while a single video covers both participants in the control groups. Additionally, four 30-minute group interviews are conducted post-experiment, resulting in 17 files and 250 minutes of Finnish-language data.

\textbf{Speech-to-Text Processing:} To automate video and audio conversion to text, we evaluate transcription tools using manually transcribed data from a previous study. Accuracy is measured with Character Error Rate (CER) \cite{morris2004character}, calculated by comparing the number of insertions, deletions, and substitutions with the ground truth. A lower CER indicates better accuracy. All files were converted to `.wav` format. Three tools were tested:

\begin{itemize}[noitemsep, topsep=0pt]
    \item \textbf{Google Cloud} \cite{googletranscribe}: Best performance with CER 0.24 overall, 0.31 for group discussions, and 0.17 for two-person conversations.
    \item \textbf{Amazon Transcription} \cite{amazontranscribe}: CER of 0.29.
    \item \textbf{Aalto ASR} \footnote{\url{https://www.kielipankki.fi/tuki/aalto-asr-automaattinen-puheentunnistin/}}: Found unsuitable due to poor recognition.
\end{itemize}

Transcriptions are processed on European servers: Aalto ASR in Finland, Amazon in Ireland, and Google in Sweden. See Appendix~\ref{appendix:cer-analysis} for detailed results.

\textbf{Text Processing and Translation:} Transcription outputs contain errors such as misrecognitions and missing punctuation. To improve quality and translate Finnish to English, we use a multi-step process (visualized in Appendix~\ref{appendix:data-pipeline}): GPT-3.5 \cite{openai2023gpt} handles grammar corrections, GPT-4 \cite{openai2023gpt} performes context-aware refinements and translation, and DeepL \cite{deepl} provides additional translation support. The English-speaking researcher uses the translated output for annotation tasks.

\textbf{Data Annotation:} After preprocessing, 8 individuals and 2 LLMs annotated speaker segments and emotions based on Plutchik’s wheel \cite{plutchik1980general}. Annotations are created using Doccano \cite{doccano}, supporting overlapping sequences.

\textbf{Annotation Guidelines and Evaluation:}
\begin{itemize}[noitemsep, topsep=0pt]
    \item \textbf{Speaker Marking:} Dialogue is segmented by speaker.
    \item \textbf{Emotion Labeling:} Sentences are labeled with emotions such as joy, sadness, anger, fear, surprise, disgust, trust, and anticipation.
\end{itemize}

Annotation agreements are evaluated using Cohen’s kappa \cite{cohen1960} (inter-rater reliability) and Jaccard similarity \cite{jaccard1901distribution} (set overlap), with metric definitions in Appendix~\ref{appendix:annotation-metrics}.

\section{Evaluation}
In the MR condition, the added spatial dimension allows for more realistic perception of depth and motion, with participants appearing co-located in the same environment. This setup is expected to support more natural collaboration and engagement. Although MR enables the study of gestures and shared spatial references, our evaluation focuses on language-based interaction.

The evaluation setup involves human annotators, language models, and a combination of both. The original transcribed text is in Finnish, and Finnish-speaking annotators receive the data directly from the transcription system. In contrast, the English-speaking annotator works with corrected data from a language model—selected based on its lower CER—which is then translated using DeepL to enhance accuracy and reliability. The goal is to compare the performance of Finnish-speaking annotators, language models (e.g., GPT-3 and GPT-4), and the English-speaking annotator utilizing AI tools.

Some annotations are completed by a single annotator (language models and the non-Finnish-speaking annotator), while multiple Finnish-speaking annotators participated. The results for Finnish speakers are reported as both the average performance (Finn-average) and the performance of the annotator with the highest agreement with other Finnish-speaker annotators (Finn-top). Data are divided into three categories: (1) control group data (Zoom), where two children play a game while an adult instructs, with transcripts labeled “Speaker 1,” “Speaker 2,” and “Instructor”; (2) experimental group data (MR), where each child's voice is captured through a dedicated microphone, eliminating the need for speaker recognition; and (3) group interviews (post-study), where five to six children interact with an adult, and transcripts are labeled as “Interviewer” and “Interviewee” due to the challenge of distinguishing individual speakers.
\subsection{Speaker Identification Task}

Speaker identification agreement was evaluated at the sentence level using Cohen's Kappa to ensure consistency across annotators and conditions. The analysis focused on two roles: “Finn-top”, representing the Finnish-speaking annotator with the highest agreement relative to others, and “Finn-average”, representing the mean agreement across all Finnish annotators and pairs. Each annotator’s labels were compared against real speaker, that we manually labeled, labels derived from the recordings. Finn-top achieved the highest agreement in both interview (\(\kappa = 0.5348\)) and experimental (\(\kappa = 0.928\)) data. The En-speaker showed lower agreement (\(\kappa = 0.1157/0.752\)), likely due to translation-related information loss, though their performance in interview data remained acceptable. Notably, GPT-4 outperformed both GPT-3.5 and the En-speaker in the experimental condition (\(\kappa = 0.4353\) vs. \(0.5995/0.752\)), whereas GPT-3.5 performed better in group interviews (\(\kappa = 0.0381\) vs. \(0.2391/0.1157\)), suggesting that GPT-3.5 may be more effective in simpler, dialogic contexts, while GPT-4 better handles more complex scenes (see Appendix~\ref{appendix:GPT-examples}). Overall, Finnish-speaking annotators demonstrated strong agreement with one another and with the real speaker references, particularly in the group interviews (\(\kappa = 0.917/0.503\)), and maintained relatively high agreement even in the more challenging experimental sessions (\(\kappa = 0.928/0.535\)) (see full breakdown in Appendix~\ref{appendix:speaker-agreement}).

\begin{figure}[h]
    \centering
    \resizebox{\columnwidth}{!}{%
        \begin{tikzpicture}
            \begin{axis}[
                width=\columnwidth,  height=7cm,  
                ybar, bar width=31pt, x=2.7cm, 
                symbolic x coords={En-speaker, Finn-top, Finn-average, GPT-4, GPT-3.5},
                xtick=data, 
                ymin=0, ymax=1, 
                ylabel={Cohen's Kappa}, 
                nodes near coords,
                enlarge x limits=0.15,  
                legend pos=north west
            ]
                \addplot coordinates {(En-speaker,0.551) (Finn-top,0.661) (Finn-average,0.651) (GPT-4,0.030) (GPT-3.5,0.017)};
                \addplot coordinates {(En-speaker,0.555) (Finn-top,0.718) (Finn-average,0.700) (GPT-4,0.040) (GPT-3.5,0.036)};
                \legend{Interviews, Experiments}
            \end{axis}
        \end{tikzpicture}
    }
    \Description{En-speaker is against real switches, Finn-top and Finn-average are simply mean}
    \caption{Annotation Agreement for Speaker Switch Task}
    \label{fig:speaker-switch-agreement}
\end{figure}
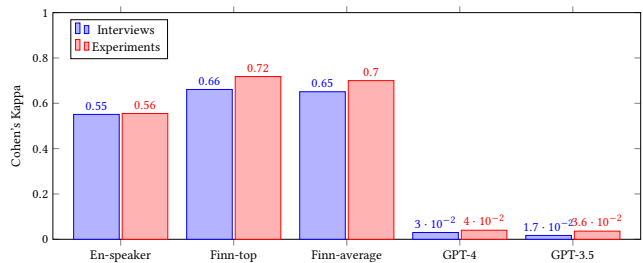
Even when the exact moment of a speaker switch is correctly identified, annotators or models may mislabel the speaker, or miss a few switches—especially in longer dialogues—leading to cascading errors that lower overall agreement. These issues are particularly evident in the experimental data, where speaker dynamics and context shifts are more complex. Figure \ref{fig:speaker-switch-agreement} illustrates this effect by showing annotator agreement on the precise point of speaker changes. While agreement is generally high for both English and Finnish, a notable gap appears between speaker identification and switch-level agreement in the experimental setting. This suggests that although annotators often detect where a switch occurs, missing just a few instances can significantly affect final scores. Transcription models also face challenges in group settings, where overlapping speech and child voices—underrepresented in training data—further complicate the task. Nonetheless, annotators may mitigate ambiguity by using generic labels like “Interviewer,” which helps preserve agreement. While Cohen’s kappa remains a standard measure, examining these nuanced errors reveals important limitations. In contrast to human annotators, LLMs show low switch-level agreement, indicating that they often over-label dominant speakers and overlook others, despite strong overall performance in speaker identification—highlighting an area for further improvement.

\subsection{Emotion Detection Task}
The evaluation of emotion detection assessed agreement between various annotators and Finn-top, with Finn-average values provided for an overall perspective. We observed that the majority of annotated emotions fall into the positive category—Anticipation, Joy, Trust, and Surprise—compared to the negative group, which includes Fear, Sadness, Anger, and Disgust. This overall distribution, highlighting the predominance of positive emotions, is further illustrated in the general visualization provided in Appendix~\ref{appendix:emotion-distribution}.

\begin{figure}[h]
\centering
    \resizebox{\columnwidth}{!}{%
\begin{tikzpicture}
    \begin{polaraxis}[
        title={Emotion Distribution},
        xtick={0, 45, 90, 135, 180, 225, 270, 315},
        xticklabels={Trust, Surprise, Fear, Sadness, Anger, Disgust, Anticipation, Joy},
        ymajorgrids=true,
        ymin=0, ymax=3000, 
        legend pos=outer north east
    ]
    \addplot[color=purple, line width=1.5pt, opacity=0.8] coordinates {
        (0,842.5) (45,462.5) (90,237.5) (135,270.0) (180,477.5) (225,212.5) (270,722.5) (315,970.0) (360,842.5)
    };
    \addlegendentry{Interviews}

    \addplot[color=teal, line width=1.5pt, opacity=0.8] coordinates {
        (0,1236.7) (45,1023.3) (90,330.0) (135,190.0) (180,266.7) (225,156.7) (270,1963.3) (315,810.0) (360,1236.7)
    };
    \addlegendentry{Zoom}

    \addplot[color=orange, line width=1.5pt] coordinates {
        (0,2106.0) (45,1352.0) (90,564.0) (135,388.0) (180,884.0) (225,398.0) (270,2548.0) (315,1230.0) (360,2106.0)
    };
    \addlegendentry{MR}
    
    \end{polaraxis}
\end{tikzpicture}%
}
\caption{Emotion Distribution by Group Data (Scaled)}
\Description{Scaled emotion distribution by group to account for different session counts.}
\label{fig:emotion-distribution-groups}
\end{figure}
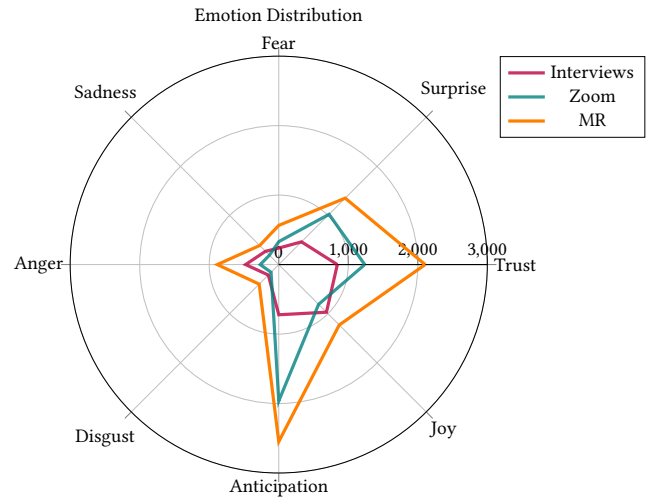

Figure~\ref{fig:emotion-distribution-groups} presents the distribution of each emotion separately across platforms, using scaled emotion counts normalized by the session count of three groups (4 interviews, 3 Zoom, and 5 MR). The Interview Group contains 71.45\% positive emotions, the MR group 76.40\% positive, and the Zoom group 84.21\% positive. The larger area under the MR curve indicates greater emotional engagement among children.


Sentiment annotation agreement results show that in both the control (Zoom) and MR datasets, GPT-3.5 consistently outperforms the average of Finnish-speaking annotators (Finn-average), indicating its effectiveness in detecting sentiment in structured, two-person dialogues (\(\kappa = 0.9306/0.7913\) vs. \(0.691/0.613\)). GPT-4 and the English-speaking annotator (En-speaker) perform at levels comparable to Finn-average across both datasets (GPT-4: \(\kappa = 0.6845/0.6016\); En-speaker: \(0.6810/0.6362\)). In contrast, the interview data present a greater challenge, with GPT-3.5 achieving its lowest agreement (\(\kappa = 0.7067\)) and Finn-average the highest (\(\kappa = 0.768\)), suggesting that human annotators handle complex, less structured interactions more reliably. Interestingly, GPT-3.5 outperforms GPT-4 overall (GPT-4 interview: \(\kappa = 0.5943\)), indicating stronger handling of conversational context and style. Finnish annotators particularly excel as the annotation task grows more complex, such as in spontaneous and overlapping group discussions (\(\kappa = 0.663\)). Additional agreement metrics are provided in Appendix~\ref{appendix:sentiment-agreement}.

\subsection{Resources}
The human annotators (one English speaker and two Finnish speakers) spent 32 hours on the task, while automated models completed it in minutes—GPT-4 in 8 minutes and GPT-3.5 in 3 minutes (see Appendix~\ref{appendix:time-breakdown}). At the time of this study, according to InfoFinland\footnote{\url{https://www.infofinland.fi/en/work-and-enterprise/during-employment/wages-and-working-hours}}, the median minimum wage in Finland is €9.16 per hour, and GPT-4-turbo costs approximately \$0.01 per 1000 characters for input and \$0.03 per 1000 characters for output\footnote{\url{https://openai.com/api/pricing/?utm_source=chatgpt.com}}. This comparison clearly demonstrates the significant time and cost efficiencies that automated tools offer over human annotators for extensive annotation tasks.

\section{Conclusion}
This study examines the impact of Mixed Reality (MR) versus traditional 2D video conferencing (Zoom) on children’s communication behaviors within an educational context. By leveraging human annotators and advanced AI models (GPT-3.5 and GPT-4), the methodology processes video and audio recordings to extract meaningful emotional insights while addressing limitations in transcription and annotation. Notably, transcription tools and LLMs performed more effectively in two-person interactions, with accuracy potentially improving further using datasets predominantly featuring adult speakers. AI tools also enabled a non-Finnish speaker to analyze the same data—yielding sentiment analysis results in two-person interactions that closely align with the average performance of Finnish-speaking annotators—thereby bridging language barriers.

The study found that all three setups—Experiment, Zoom, and MR—successfully fostered positive engagement among children. However, while MR offers immersive experiences, it did not surpass Zoom in promoting positive engagement. Technical challenges and unresolved bugs in the MR system may have affected children's emotional responses and overall experience, leading to occasional disruptions or frustrations. Despite these issues, the findings suggest that MR remains a promising tool for enhancing children's interactions, with potential for improvement as the technology evolves.

\section{Future Research}
 Further investigation is needed to refine our understanding of presence and immersion in Mixed Reality (MR) environments, with contributions from psychology to better define and measure these concepts. Future research will also extend the current workflow—from audio transcription to analysis—by applying it to a new dataset with an improved setup that incorporates adult participants and better technical conditions for higher-quality results. Additionally, with the emergence of multimodal GPT models capable of processing both audio and text, their potential for transcription and analysis will be explored and compared to existing methods. Ongoing projects will continue to assess the role of AI tools in analyzing sentiment and immersion, while future MR technology development will focus on optimizing live holograph-based holoportation solutions, addressing current limitations associated with avatar-based MR that may offer a lesser sense of presence \cite{riedl2011trusting}.

\begin{acks}

This publication's data has been produced under the financing of the Research Council of Finland.

We sincerely thank Calkin Suero Montero, Eero Nirhamo, Sebastian Hahta, who contributed to this study by assisting with data collection, annotation, and the selection of a psychological model.
\end{acks}

\bibliographystyle{ACM-Reference-Format}
\bibliography{draft-references}


\appendix
\section{Tables}

\subsection{Character Error Rate (CER) Analysis}
\label{appendix:cer-analysis}

This appendix provides detailed CER measurements across transcription tools, conversation types, and LLM-based correction methods.

\begin{table}[H]
\centering

\begin{minipage}{0.48\textwidth}
\centering
\caption{CER by Speech-to-Text Tool}
\begin{tabular}{lc}
\hline
\textbf{Tool} & \textbf{CER} \\ \hline
Amazon Transcription & 0.29 \\ \hline
Google Cloud         & 0.24 \\ \hline
\end{tabular}
\label{tab:cer-table}
\end{minipage}
\hfill
\begin{minipage}{0.48\textwidth}
\centering
\caption{Google Cloud CER by Data Type}
\begin{tabular}{lc}
\hline
\textbf{Data Type} & \textbf{CER} \\ \hline
Group Discussions         & 0.31  \\ \hline
Two-Person Conversations  & 0.175 \\ \hline
\end{tabular}
\label{tab:cer-details}
\end{minipage}

\end{table}

\vspace{1em}

\begin{table}[H]
\centering
\caption{CER After Corrections with LLMs}
\begin{tabular}{lc}
\hline
\textbf{Model} & \textbf{CER} \\ \hline
GPT-3.5        & 0.180         \\ \hline
GPT-4          & 0.178         \\ \hline
\end{tabular}
\label{tab:cer-Refinement}
\end{table}

\subsection{Annotation Time Breakdown}
\label{appendix:time-breakdown}
Table~\ref{tab:annotation-resources} shows the time each annotator or model spent completing the annotation task. This highlights the stark contrast in efficiency between human annotators and automated models.

\begin{table}[H]
\caption{Time Spent by Each Annotator}
\centering
\begin{tabular}{lc}
\hline
\textbf{Annotator} & \textbf{Time spent}    \\ \hline
English Speaker          & 16 hours         \\ \hline
Finnish Speakers         & 8 hours          \\ \hline
GPT-4                    & 8 minutes        \\ \hline
GPT-3.5                  & 3 minutes        \\ \hline
\end{tabular}
\label{tab:annotation-resources}
\end{table}

\subsection{Speaker Identification Annotation Agreement}
\label{appendix:speaker-agreement}

This appendix provides Cohen’s Kappa metrics for speaker identification. Table~\ref{tab:speaker-annotation-agreement} shows agreement with real speaker labels, while Table~\ref{tab:speaker-annotation-agreement-Finn} summarizes agreement among Finnish-speaking annotators.

\begin{table}[h]
\centering
\begin{minipage}{0.48\textwidth}
\centering
\caption{Agreement with Real Speakers}
\begin{tabular}{lcc}
\hline
\textbf{Annotator} & \textbf{Kappa (Exp)} & \textbf{Kappa (Interviews)} \\ \hline
Finn-top       & 0.928 & 0.5348 \\
Finn-average   & 0.917 & 0.5028 \\
En-speaker     & 0.752 & 0.1157 \\
Finn-top       & 0.941 & 0.3410 \\
GPT-4          & 0.4353 & 0.2391 \\
GPT-3.5        & 0.5995 & 0.0381 \\ \hline
\end{tabular}
\label{tab:speaker-annotation-agreement}
\end{minipage}
\hfill
\begin{minipage}{0.48\textwidth}
\centering
\caption{Agreement Among Finnish Annotators}
\begin{tabular}{lcc}
\hline
\textbf{Annotator} & \textbf{Kappa (Exp)} & \textbf{Kappa (Interviews)} \\ \hline
Finn-top       & 0.5348 & 0.928 \\
Finn-average   & 0.5028 & 0.917 \\ \hline
\end{tabular}
\label{tab:speaker-annotation-agreement-Finn}
\end{minipage}
\end{table}

\subsection{Sentiment Annotation Agreement}
\label{appendix:sentiment-agreement}

To support the sentiment agreement analysis in the main text, we present Cohen's kappa scores for each dataset in the following tables.

\begin{table}[H]
\centering

\begin{minipage}{0.48\textwidth}
\centering
\caption{Sentiment Agreement – MR Platform}
\begin{tabular}{llc}
\hline
\textbf{Annotator 1} & \textbf{Annotator 2} & \textbf{Cohen's kappa} \\ \hline
Finn-top             & -                    & 0.696                  \\ \hline
Finn-average         & -                    & 0.613                  \\ \hline
En-speaker           & Finn-top             & 0.6362                 \\ \hline
GPT-4                & Finn-top             & 0.6016                 \\ \hline
GPT-3.5              & Finn-top             & 0.7913                 \\ \hline
\end{tabular}
\label{tab:sentiment-annotation-agreement-MR}
\end{minipage}
\hfill
\begin{minipage}{0.48\textwidth}
\centering
\caption{Sentiment Agreement – Control Group (Zoom)}
\begin{tabular}{llc}
\hline
\textbf{Annotator 1} & \textbf{Annotator 2} & \textbf{Cohen's kappa} \\ \hline
Finn-top             & -                    & 0.765                  \\ \hline
Finn-average         & -                    & 0.691                  \\ \hline
En-speaker           & Finn-top             & 0.6810                 \\ \hline
GPT-4                & Finn-top             & 0.6845                 \\ \hline
GPT-3.5              & Finn-top             & 0.9306                 \\ \hline
\end{tabular}
\label{tab:sentiment-annotation-agreement-zoom}
\end{minipage}

\end{table}

\begin{table}[H]
\caption{Sentiment Agreement – Interview Dataset}
\centering
\begin{tabular}{llc}
\hline
\textbf{Annotator 1} & \textbf{Annotator 2} & \textbf{Cohen's kappa} \\ \hline
Finn-top             & -                    & 0.663                  \\ \hline
Finn-average         & -                    & 0.768                  \\ \hline
En-speaker           & Finn-top             & 0.56                   \\ \hline
GPT-4                & Finn-top             & 0.5943                 \\ \hline
GPT-3.5              & Finn-top             & 0.7067                 \\ \hline
\end{tabular}
\label{tab:sentiment-annotation-agreement-interviews}
\end{table}

\section{Figures}

\subsection{Case Study}
A picture from case study 2, with rooms setups.

\begin{figure}[H]
  \includegraphics[width=\linewidth]{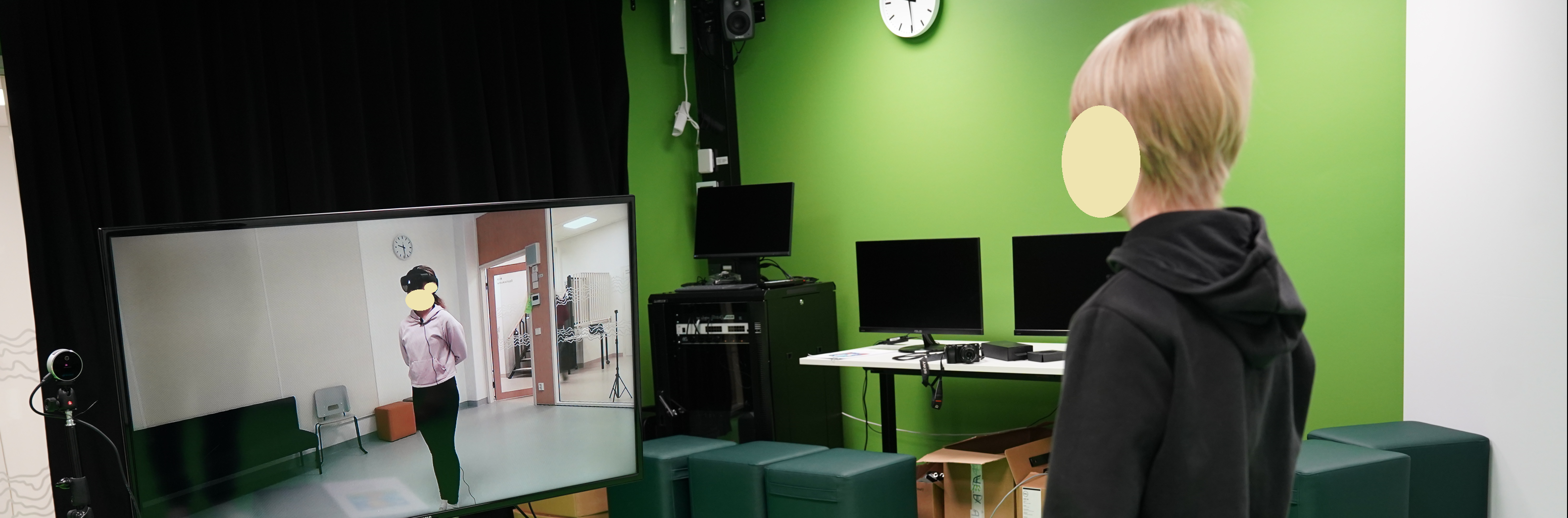}
  \caption{Case Study Setup, Dec 2023.}
  \Description{A pair of children interacting through a MR headset and a 2D monitor.}
  \label{fig:teaser}
\end{figure}

\subsection{Data Processing Pipeline}
\label{appendix:data-pipeline}

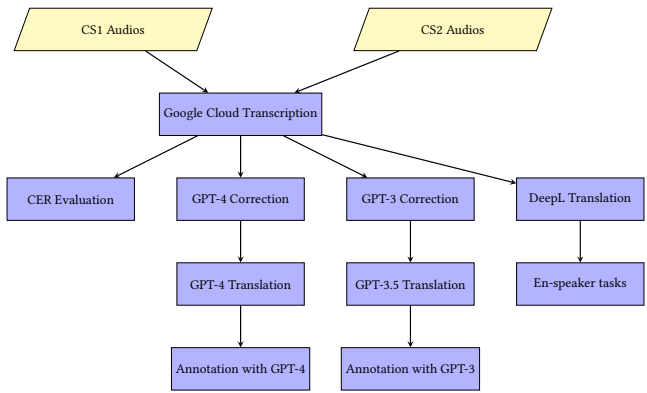
\begin{figure}[H]
    \centering
\resizebox{\columnwidth}{!}{%
\begin{tikzpicture}[node distance=2cm]

\node (cs1audio) [data] {CS1 Audios};
\node (cs2audio) [data, right of=cs1audio, xshift=6cm] {CS2 Audios};
\node (transcription) [process, below of=cs1audio, xshift=3cm] {Google Cloud Transcription};
\node (cer) [process, below of=transcription, xshift=-4cm] {CER Evaluation};
\node (gpt4) [process, below of=transcription] {GPT-4 Correction};
\node (gpt3) [process, below of=transcription, xshift=4cm] {GPT-3 Correction};
\node (deepl) [process, right of=gpt3, xshift=2cm] {DeepL Translation};
\node (gpt4translation) [process, below of=gpt4] {GPT-4 Translation};
\node (gpt3translation) [process, below of=gpt3] {GPT-3.5 Translation};
\node (deepltranslation) [process, right of=gpt3translation, xshift=2cm] {En-speaker tasks};
\node (gpt4marking) [process, below of=gpt4translation] {Annotation with GPT-4};
\node (gpt3marking) [process, below of=gpt3translation] {Annotation with GPT-3};

\draw [arrow] (cs1audio) -- (transcription);
\draw [arrow] (cs2audio) -- (transcription);
\draw [arrow] (transcription) -- (cer);
\draw [arrow] (transcription) -- (gpt4);
\draw [arrow] (transcription) -- (gpt3);
\draw [arrow] (transcription) -- (deepl);
\draw [arrow] (gpt4) -- (gpt4translation);
\draw [arrow] (gpt3) -- (gpt3translation);
\draw [arrow] (deepl) -- (deepltranslation);
\draw [arrow] (gpt4translation) -- (gpt4marking);
\draw [arrow] (gpt3translation) -- (gpt3marking);

\end{tikzpicture}%
}   
\caption{Data Flow – From Recordings to Annotations}
\Description{Data pipeline for both methodology and evaluation process}
\label{fig:data-pipeline}
\end{figure}

Figure~\ref{fig:data-pipeline} provides an overview of the multi-step data flow used in our annotation pipeline, including transcription, correction, translation, and automated annotation using large language models.

\subsection{Emotion Distribution Visualization}
\label{appendix:emotion-distribution}

To complement our discussion on emotion annotation, Figure~\ref{fig:positive-negative-emotions} provides a scaled visualization of emotion distributions across Interview, Zoom, and MR datasets. This presentation allows for an aggregated perspective across different modalities while accounting for session count differences.

\definecolor{negative}{HTML}{FDAE61}
\definecolor{positive}{HTML}{ABDDA4}
\begin{figure}[H]
\centering
\resizebox{\columnwidth}{!}{%
\begin{tikzpicture}[x={(.01,0)}]
\foreach  \l/\x/\c[count=\y] in {
negative emotions/705/negative,  
positive emotions/2244/positive} 
{\node[left] at (0,\y) {\l};
\fill[\c] (0,\y-.4) rectangle (0.26 * \x,\y+.4);
\node[right] at (0.26 *\x, \y) {\x};}
\draw (0,0) -- (600,0);
\foreach \x in {400, 800, ..., 2200}
{\draw (0.26 * \x,.2) -- (0.26 * \x,0) node[below] {\x};}
\draw (0,0) -- (0,4.5);
\end{tikzpicture}%
}
\caption{Positive and Negative Emotions Distribution}
\Description{sentiment}
\label{fig:positive-negative-emotions}
\end{figure}
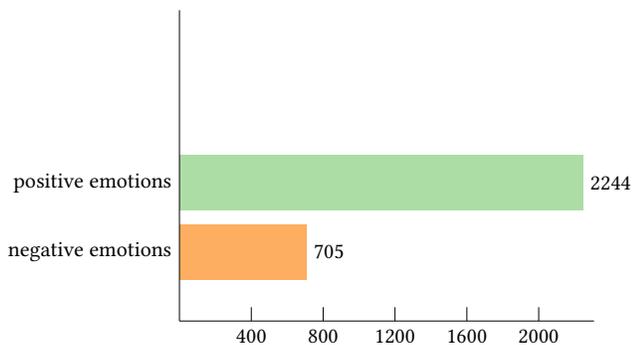

\section{Formulas}
\subsection{Annotation Metrics}
\label{appendix:annotation-metrics}

This appendix provides the formal definitions of the metrics used to assess annotation agreement and similarity.

\begin{itemize}
    \item \textbf{Cohen’s kappa:}
    \[
    \kappa = \frac{P_o - P_e}{1 - P_e}
    \]
    \begin{itemize}
        \item $P_o$ is the observed agreement among raters.
        \item $P_e$ is the expected agreement (chance agreement).
    \end{itemize}

    \item \textbf{Jaccard similarity:}
    \[
    J(A, B) = \frac{|A \cap B|}{|A \cup B|}
    \]
    \begin{itemize}
        \item $|A \cap B|$ is the size of the intersection of sets $A$ and $B$.
        \item $|A \cup B|$ is the size of the union of sets $A$ and $B$.
    \end{itemize}
\end{itemize}

\section{GPT Annotation Examples}
\label{appendix:GPT-examples}

\subsection*{Example: Emotion Annotation Comparison}

\textbf{Transcript Excerpt:}
\begin{quote}
\emph{"And indeed, I was afraid at first, but then I thought it might be fun, and I smiled. Still, I didn’t know what would happen next."}
\end{quote}

\textbf{GPT-3 and GPT-4 Emotion Annotations (4 emotions):}
\begin{itemize}
    \item Joy
    \item Fear
    \item Anticipation
    \item Trust
\end{itemize}

With such a complex sentence, GPT-4 provides a more nuanced emotional annotation than GPT-3. It disaggregates the emotional trajectory across the sentence, placing multiple distinct segments (especially for anticipation).

\appendix

\end{document}